\documentclass[11pt]{revtex4}    % Specifies the document style.
\usepackage{amssymb,epsf}
\usepackage{latexsym}
\begin{document}

\title{Asymptotically non-flat Einstein-Born-Infeld-dilaton
 black holes with Liouville-type potential}
\author{A. Sheykhi, N. Riazi\footnote{email address:
riazi@physics.susc.ac.ir}, and M. H. Mahzoon}
\address{Physics Department and Biruni Observatory, College of Sciences,
Shiraz University, Shiraz 71454, Iran}

\begin{abstract}
We construct some classes of electrically  charged, static and
spherically symmetric black hole solutions of the four-dimensional
Einstein-Born-Infeld-dilaton gravity in the absence and  presence
of Liouville-type potential for the dilaton field and investigate
their properties. These solutions are neither asymptotically flat
nor (anti)-de Sitter. We show that in the presence of the
Liouville-type potential, there exist two classes of solutions. We
also compute temperature, entropy, charge and mass of the black
hole solutions, and find that these quantities satisfy the first
law of thermodynamics. We find that in order to fully satisfy all
the field equations consistently, there must be a relation between
the electric charge and other parameters of the system.

\end{abstract}

\maketitle
\section{Introduction}
It is quite possible that gravity is not given by the Einstein
action, at least at sufficiently high energies. In string theory,
gravity becomes scalar-tensor in nature. The low energy limit of
the string theory leads to the Einstein gravity, coupled
non-minimally to a scalar dilaton field \cite{Wit1}. When a
dilaton is coupled to Einstein-Maxwell theory, it has profound
consequences for the black hole solutions. Some efforts have been
done to construct exact solutions of Einstein-Maxwell-dilaton
gravity. For example exact charged dilaton black hole solutions of
EMd gravity in the absence of a dilaton potential have been
constructed by many authors \cite{CDB1,CDB2}. The dilaton changes
the casual structure of the spacetime and leads to curvature
singularities at finite radii. These black holes are
asymptotically flat. In recent years, non-asymptotically flat
black hole spacetimes are attracting much interest in connection
with the so called AdS/CFT correspondence. Black hole spacetimes
which are neither asymptotically flat nor dS/AdS have been found
and investigated by many authors. The uncharged solutions have
been found in \cite{MW}, while the charged solutions have been
considered in \cite{PW}. In the presence of Liouville-type
potential, static charged black hole solutions have also been
discovered with a positive constant curvature event horizons and
zero or negative constant curvature horizons \cite{CHM,Cai1}. The
extension to the dyonic black hole solutions
 in four-dimensional and higher dimensional EMd gravity with one and
 two Liouville-type potentials  have also been done in \cite{yaz1}. These solutions
possess both electric and magnetic charge and they are neither
asymptotically flat nor dS/AdS.

The idea of the nonlinear electromagnetism was first introduced in
1934 by Born and Infeld in order to obtain a finite value for the
self-energy of point-like charges\cite{BI}. Although it become
less popular with the introduction of  \textit{QED}, in recent
years, the Born-Infeld action has been occurring repeatedly with
the development of superstring theory, where the dynamics of
D-branes is governed by the Born-Infeld action \cite{Fra,Cal}. For
various motivations, extending the Reissner- Nordstr\"{o}m black
hole solutions in EM theory to the charged black hole solutions in
EBI theory has attracted some attention in recent years\cite{wil}.
For example, exact solutions of spherically symmetric
Einstein-Born-Infeld black holes in (A)dS spacetime with
cosmological horizon in arbitrary dimensions has been  constructed
in \cite{dey}. The extension to the case where black hole horizon
(cosmological horizon) is a positive, zero or negative constant
curvature surface have also been studied \cite{Cai2}.
Unfortunately,  exact  solutions to the Einstein-Born-Infeld
equation coupled to matter fields are  too complicated to find
except in a limited number of cases. Indeed, exact solutions to
the Einstein Born-Infeld dilaton (EBId) gravity are known only in
three dimensions \cite{YI}. Numerical studies of the EBId system
in four dimensional static and spherically symmetric spacetime
have been done in \cite{yaz2}. In  the absence of a dilaton
potential a class of solution to the four-dimensional EBId gravity
with magnetic charge has been constructed \cite{yaz3}. Our aim in
this paper is to  generalize these solutions to the case of one
and two Liouville type potential and investigate how the
properties of the solutions will be changed in the presence of
potential for the scalar field. In addition, in each cases, we
compute, the mass, electric charge, temperature and entropy of the
system. We will consider three special cases: (a) $V(\phi)=0$, (b)
$V(\phi)={2\Lambda}e^{2\beta\phi}$ and (c) $V(\phi)=2\Lambda_1
e^{2\beta_{1}\phi} +2\Lambda_2 e^{2\beta_{2}\phi}$. The first case
corresponds to the action considered in \cite{gib}. When
$\alpha=1$, it reduces to the four-dimensional low-energy action
obtained from string theory in terms of Einstein metric. Case (b)
corresponds to a Liouville-type potential. This kind of potential
appear when one applies  a conformal transformation on the low
energy limit of the string tree level effective action for
massless boson sector and write the action in the Einstein frame
\cite{gross,deh1}. This potential have been considered previously
by a number of authors\cite{CHM,yaz1,deh2}. One may refer to
$\Lambda $ as the cosmological constant, since in the absence of
the dilaton field the action  reduces to the action of EBI gravity
with cosmological constant \cite{dey,Cai2}. The potential in case
(c) was previously investigated by a number of authors both
 in the context of $FRW$ scalar field cosmologies \cite{ozer} and
EMd black holes \cite{CHM,yaz1}. This kind of potential function
can be obtained when a higher-dimensional theory is compactified
to four-dimensional spacetime, including various supergravity and
string models.

The organization of this paper is as follows: Section \ref{Field}
is devoted to a brief review of the field equations and general
equations of motion. In Sec. \ref{Vzero} we consider EBId black
holes without potential. In Sec. \ref{V}, we present two classes
of solutions with a Liouville type potential and general dilaton
coupling. In Sec. \ref{2V}, we extend these solutions to the case
of two Liouville potentials. We finish our paper with some
concluding remarks.

\section{Field Equations}\label{Field}
We consider the four-dimensional action in which gravity is
coupled to  dilaton and Born Infeld  field with an action
\begin{equation}\label{action}
 S = \int d^4x\sqrt{-g}\left({\cal R}
-2(\nabla{\phi})^2 - V(\phi)+ L(F,\phi)\right)
\end{equation}
where ${\cal R}$ is the Ricci scalar curvature, $\phi$ is the
dilaton field and $V(\phi)$ is a potential for $\phi$. The
Born-Infeld $L(F,\phi)$ part of the action is given by
\begin{equation}\label{BI1}
L(F,\phi)= 4\gamma e^{-2\alpha\phi} \left( 1 -\sqrt{1 +
\frac{F^{\mu\nu} F_{\mu\nu}}{2\gamma}}\right).
\end{equation}
Here, $\alpha$ is the dilaton coupling constant and $ \gamma $ is
called the Born-Infeld parameter with dimension of mass. In the
limit $\gamma\rightarrow \infty$, $L(F,\phi)$ reduces to the
standard Maxwell field coupled to a dilaton field
\begin{equation}\label{BI2}
L(F,\phi)=- e^{-2\alpha\phi}{F^{\mu\nu}F_{\mu\nu}}.
\end{equation}
On the other hand, $L(F,\phi)\rightarrow 0$ as $\gamma
\rightarrow0$. It is convenient to set
\begin{equation}\label{BI3}
L(F,\phi)= 4\gamma e^{-2\alpha\phi}{\cal{L}}(Y).
\end{equation}
where
\begin{eqnarray}
{\cal{L}}(Y)&=&1-\sqrt{1 + Y},\label{LY} \\
&&Y=\frac{F^{2}}{2\gamma}\label{Y}.
\end{eqnarray}
where $F^{2}={F^{\mu\nu}F_{\mu\nu}}$. The equations of motion can
be obtained by varying the action (\ref{action}) with respect to
the gravitational field $g_{\mu \nu }$, the dilaton field $\phi $
and the gauge field $A_{\mu }$ which yields the following field
equations
\begin{equation}\label{FE1}
{\cal R}_{\mu\nu} = 2\partial_{\mu}\phi\partial_{\nu}\phi
+\frac{1}{2}g_{\mu\nu}V(\phi)- 4e^{-2\alpha\phi}\partial_{Y}{{\cal
L}}(Y) F_{\mu\eta} F_{\nu}^{\text{ }\eta }+2\gamma
e^{-2\alpha\phi} \left[2Y\partial_{Y}{{\cal L}}(Y)-{{\cal
L}}(Y)\right]g_{\mu\nu},
\end{equation}
\begin{equation}\label{FE2}
\nabla ^{2}\phi =\frac{1}{4}\frac{\partial V}{\partial \phi}+
2\gamma\alpha e^{-2\alpha\phi}{\cal L}(Y),
\end{equation}
\begin{equation}\label{FE3}
\nabla_{\mu}\left(e^{-2\alpha\phi} \partial_{Y}{{\cal L}}(Y)
F^{\mu\nu}\right)=0.
\end{equation}

We wish to find static and spherically symmetric solutions of the
above field equations. The most general such metric can be written
in the form
\begin{equation}\label{metric}
ds^2=-U(r)dt^2 + {dr^2\over U(r)} + R^2(r)\left(d\theta^2 +
\sin^2\theta d\varphi^2 \right).
\end{equation}

The Maxwell equation can be integrated immediately, where all the
components of $F^{\mu\nu}$ are zero except $ F^{rt}$:
\begin{equation}\label{Frt}
F^{rt}= \frac{q e^{2\alpha\phi}}{ R^2(r)\sqrt {1+{\frac
{q^{2}e^{4\alpha\phi}}{\gamma R^4(r)}}}}
\end{equation}
Here $q$ is the electric charge,  defined through the integral
\begin{equation}\label{Q}
q = \frac{1}{4\pi}\int_{s^2} e^ {-2\alpha\phi} *F d{\Omega},
\end{equation}
where $*$ is the Hodge dual and $ s^{2}$ is any two-sphere defined
at spatial infinity, which its volume element denoted by
$d{\Omega}$. We note that the electric field is finite at $r=0$.
This is expected in Born-Infeld theories. It is interesting to
consider three limits of (\ref{Frt}). First, for large $\gamma$
(where the BI action reduces to Maxwell case) we have
$F^{rt}=\frac{q e^{2\alpha\phi}}{R^2(r)}$  as presented in
\cite{CHM}. On the other hand, if $\gamma\rightarrow 0$ we get
$F^{rt}=0$, finally in the case of $\alpha \rightarrow 0$ it
reduces to the case of Einstein Born Infeld theory without dilaton
field \cite{dey,Cai2}. With the metric (\ref{metric}) and Maxwell
field (\ref{Frt}), the field equations (\ref{FE1}) and (\ref{FE2})
reduce to the following system of coupled ordinary differential
equations
\begin{equation}\label{ODE1}
{1\over R^2}{d\over dr }\left(U {dR^2\over dr } \right)=
\frac{2}{R^2}-V(\phi)- 4\gamma
e^{-2\alpha\phi}\left[2Y\partial_{Y}{{\cal L}}(Y)- {{\cal
L}}(Y)\right],
\end{equation}
\begin{equation}\label{ODE2}
\frac{1}{R^2}{d\over dr }\left(R^2 U {d\phi\over dr } \right)
=\frac{1}{4}\frac{d{V}}{d\phi}+2\gamma \alpha e^{-2\alpha\phi}
{\cal L}(Y),
\end{equation}
\begin{equation}\label{ODE3}
{1\over R}{d^2R\over dr^2} + \left({d\phi\over dr }\right)^2=0.
\end{equation}

In particular, in the case of the linear electrodynamics with
${\cal L}(Y)=-{1\over 2}Y$, the system of equations
(\ref{FE1})-(\ref{FE3}) and (\ref{ODE1})-(\ref{ODE3}) reduce to
the well-known  equations of EMd gravity \cite{CHM}.

To solve these equations, we make the ansatz
\begin{equation}\label{Ransatz}
R(r)=e^{\alpha \phi(r)},
\end{equation}
By this ansatz, eq. (\ref{Y}) becomes $Y=-\frac{q^2}{q^2+\gamma}$.
In addition, we introduce the constant $A$:
\begin{equation}\label{AA}
A =\gamma\left[2Y\partial_{Y}{{\cal L}}(Y)-{{\cal
L}}(Y)\right]=\sqrt{\gamma(q^2+\gamma)}-\gamma.
\end{equation}

Using (\ref{Ransatz}) in equation (\ref{ODE3}), immediately gives
\begin{equation}\label{phi}
\phi(r)=\frac{\alpha}{1+\alpha^2}\ln(br-c),
\end{equation}
where $b$ and $c$ are integration constants. For later
convenience, without loss of generality, we set $b=1$ and $c=0$.
%%%%%%%%%%%%%%%%%%%%%%%%%%%%%%%%%%%%%%%%%%%%%%%%%%%%%
\section{Solutions With $ V(\phi)=0$}\label{Vzero}
Let us begin by looking for the solutions without Liouville
potential $(V(\phi)=0)$.
\subsection{String coupling case $\alpha=1$}\label{A1}
We first consider the string coupling case $\alpha=1$ with
$V(\phi)=0$. In this case we find the following solution
\begin{equation}\label{U1}
U(r)= 2r(1-2A-\frac{r_{0}}{2r}),
\end{equation}
with $A$ is a constant related to the electric charge $q$ as one
can see from eq. (\ref{AA}) and $r_{0}>0$ is an integration
constant related to the mass of the system. In order to fully
satisfy the system of equations, there must be a relation between
electric charge $q$ and $\gamma$ parameter
\begin{equation}\label{q1}
q^2=\frac{1+\sqrt{16\gamma^2+1}}{8\gamma}
\end{equation}

Now we compute the mass of the system. In order to define the
mass, we use the so-called quasilocal formalism \cite{BY}. The
quasilocal mass is given by
\begin{equation}\label{QLM}
M = {1\over 2} {dR^2(r)\over dr} U^{1/2}(r)\left[U_{0}^{1/2}(r)
-U^{1/2}(r) \right],
\end{equation}
where $U_{0}(r)$ is an arbitrary non-negative function which
determines the zero of the energy for a background spacetime and
$r$ is the radius of the spacelike hypersurface boundary. If no
cosmological horizon is present, the large $r$ limit of
(\ref{QLM}) determines the asymptotic  mass $M$. For the solution
under consideration, there is no cosmological horizon and the
natural choice for the background is $U_{0}(r)=2(1-2A)r$. The
large $r$ limit of (\ref{QLM}) gives the mass of the solution
\begin{equation}
M = {r_{0}\over 4}.
\end{equation}

The metric corresponding to (\ref{U1}) and the other metric that
we will present in this paper are neither  asymptotically flat nor
(anti)-de Sitter. The solution has several properties. First,
there is an event horizon at $ r_{h}=2M/(1-2A)$. In order to study
the general structure of these solutions, we first look for the
curvature singularities in the presence of dilaton gravity. It is
easy to show that the Kretschmann scalar $R_{\mu \nu \lambda
\kappa }R^{\mu \nu \lambda \kappa }$ diverges at $r=0$, it is
finite for $r\neq 0$ and goes to zero as $r\rightarrow \infty $.
Also, it is notable to mention that the Ricci scaler is finite
every where except at $r=0$, and goes to zero as $r\rightarrow
\infty $. Therefore $r=r_{h}$ is a regular horizon and we have an
essential singularity located at $r=0$. This can be seen from the
explicit expression of ${\cal K}$ and ${\cal R}$:
\begin{eqnarray}\label{KR}
{\cal K}& =&\frac{1}{4r^{4}}\left[4
r^{2}\left(3-4A+12A^2\right)+4rr_{0}(1+2A)+3r_{0}^2\right],\\
&&{\cal R}=\frac{2r(6A-1)-r_{0}}{2r^{2}}.
\end{eqnarray}

Second, even though $\phi(r)$ diverges at $r\rightarrow \infty$,
but since the mass, charge and curvature all remain finite, the
solution is well behaved at infinity. Note that the dilaton field
is regular on the horizon, too. The spacial infinity is
conformally null and the solution describes a black hole with the
same causal structure as the Schwarzschild spacetime. Black hole
entropy typically satisfies the so called area law of the entropy
\cite {Beck}, which states that the entropy is a quarter of the
event horizon area. It is easy to see that the temperature  and
entropy of the black hole can be written as
\begin{eqnarray}
T &=& {1\over 4\pi} {dU \over dr}(r_{h}) = \frac{1-2A}{2\pi}, \\
S &=& \pi {r_{h}=\frac{2\pi M}{1-2A} }.\label{TS1}
\end{eqnarray}
Note that the temperature depends on $A$ and  is independent of
$M$. In the limit $A \rightarrow 1/2$, the temperature goes to
zero, while the entropy $S$ becomes infinite. Since the
temperature is always non negative quantity  thus $A\leq1/2$.

Finally we investigate the first law of thermodynamics. The black
hole solution we found here have mass and charge, thus in general
all thermodynamic quantities are functions of $M$ and $q$. Since
the electric charge is fixed, thus the first law of thermodynamics
may be written as
\begin{equation}\label{Flaw}
dM = TdS.
\end{equation}
Then it is easy to see that thermodynamics quantities obtained
above satisfy the first law (\ref{Flaw}). The solution with zero
mass ($M=0$) is singular with a null singularity at $r=0$. The
case $M<0$ corresponds to naked timelike singularity located at
$r=0$.
\subsection{General dilaton coupling $\alpha$ }\label{A2}
It is straightforward to generalize the solution (\ref{U1}) to
arbitrary dilaton coupling constant $\alpha$. In this case, we
find the following solution
\begin{equation}\label{U2}
U(r)= r^{2-2N}\left((1-2A)/N-\frac{r_{0}}{r}\right),
\end{equation}
with $N=\alpha^2/(1+\alpha^2)$. Here $r_{0}>0$ is again an
integration constant related to the mass of the black hole. The
consistency of all field equations force that, the electric charge
$q$ satisfy in the following equation:
\begin{equation}\label{q2}
\sqrt{\gamma+q^2}\left[2\gamma(\alpha^2-1)-1\right]+
2\sqrt{\gamma}\left(q^2-\gamma(\alpha^2-1)\right)=0.
\end{equation}

Note that the solution is ill defined for $\alpha=0$. In the
particular case $\alpha=1$, the solution reduces to the
(\ref{U1}). There is no cosmological horizon and the mass can be
computed from (\ref{QLM}). For the background function $U_{0}(r) =
r^{2-2N}(1-2A)/N$, the mass is found to be
\begin{equation}\label{mass}
M = \frac{N r_{0}}{2}.
\end{equation}

On the other hand, if $\alpha$ and $\gamma$ go to infinity, we
will have $N=1$ and $A=0$, respectively. Thus the metric reduces
to the Schwarzschild black hole. There is an event horizon at
$r_{h}=2M/(1-2A)$ which is regular only for $A<1/2$. The
Kretschmann invariant and Ricci scalar are finite at $r=r_{h}$,
diverge at $r=0$, and both of them vanish at $r\rightarrow
\infty$, therefore $r=r_{h}$ is a regular horizon and we have an
essential singularity located at $r=0$. Note that the dilaton
field is regular on the horizon, too. The temperature and the
entropy of the black hole on the event horizon are
\begin{eqnarray}
T &=& {1\over 4\pi} {dU \over dr}(r_{h})=\frac{1-2A}{4\pi N}r_{h}^{1-2N}, \\
S &=& \pi {r_{h}^{2N}}.
\end{eqnarray}

Finally we investigate the first law of thermodynamics. Again,
since the electric charge is fixed, thus the first law of
thermodynamics can be written as
\begin{equation}\label{Flaw2}
dM = TdS.
\end{equation}

The solution with zero mass $(M=0)$ is singular with a null
singularity at $r=0$.
%%%%%%%%%%%%%%%%%%%%%%%%%%%%%%%%%%%%%%%%%%%%%%%%%%%%%%%%%%%%%%%%%%%%%%%
\section{Solution With a Liouville type potential }\label{V}
In this section, we consider the action (\ref{action}) with a
Liouville type potential,
\begin{equation}\label{v1}
V(\phi)=2\Lambda e^{2\beta\phi},
\end{equation}
where $\Lambda$ and $\beta$ are constants. One may refer to
$\Lambda $ as the cosmological constant, since in the absence of
the dilaton field the action  reduces to the action of EBI gravity
with cosmological constant \cite{dey,Cai2}.
\subsection{Solution with $ \alpha =1$}\label{B1}
At first, we consider the case $\alpha=1$. We have found the
following solution
\begin{equation}\label{U3}
U(r)= 2r(1-2A-\Lambda - \frac{r_{0}}{2r}),
\end{equation}
with $\beta=-1$ and $r_{0}=4M$, where $M$ is the mass of the
system define via the eq. (\ref{QLM}). The electric charge, is
given by eq. (\ref{q1}). For the background function
$U_{0}(r)=(1-2A-\Lambda)r$. There is an event horizon at $
r_{h}=\frac{2 M}{1-2A-\Lambda} $ which is regular only for
$\Lambda<1-2A$. The Kretschmann invariant and Ricci scalar, are
regular except for $r=0$, where they diverge. Note that the
dilaton field is regular on the horizon, too. As an illustration
we present the Ricci scalar ${\cal R}$:
\begin{equation}\label{Ricc}
{\cal R}=\frac{2r(6A-1+3\Lambda)- r_{0}}{2r^{2}},
\end{equation}

The temperature and the entropy  of the black hole on the event
horizon are
\begin{eqnarray}
T &=& \frac{1-2A-\Lambda}{2\pi},\\
S &=& \frac{2\pi M}{1-2A-\Lambda}.
\end{eqnarray}
which satisfy  the first law. Note that the temperature depends on
$A$ and $\Lambda$ and is independent of $M$. In the limit $\Lambda
\rightarrow (1-2A)$, the temperature goes to zero, while the
entropy $S$ become infinite. It is interesting to see that our
solutions are well behaved in the limit $\Lambda \rightarrow 0$.
In other words  all of our results presented in this section
reduce to the ones presented in section (\ref{A1}), in this limit.

\subsection{Solutions with general coupling $\alpha$}
In this section, we present exact black hole solutions of EBId
with an arbitrary dilaton coupling $\alpha$ and Liouville
potential $V(\phi)=2\Lambda e^{2\beta\phi}$. In this case we can
distinguish two classes of solutions which satisfy all the field
equations depending of the suitable choice of the $\beta$
parameter.

I. $\beta=-\alpha$. In this case, using (\ref{Ransatz}) and
(\ref{phi}), one can easily show that eqs. (\ref{ODE1}) and
(\ref{ODE2}) have solution of the form
\begin{equation}\label{U4}
U(r)= \frac{r^{2-2N}}{N}\left(1-2A-\Lambda-\frac{2M}{r}\right),
\end{equation}
where $M$ is the mass of the system define via the eq.
(\ref{QLM}). Again, the consistency of all field equations force
that, the electric charge $q$ satisfy in the following equation:
\begin{equation}\label{q3}
\sqrt{\gamma+q^2}\left[(2\gamma-\Lambda)(\alpha^2-1)-1\right]+
2\sqrt{\gamma}\left(q^2-\gamma(\alpha^2-1)\right)=0.
\end{equation}

For the background function $U_{0}(r) = r^{2-2N}(1-2A-\Lambda)/N$.
There is an event horizon at $ r_{h}=\frac{2 M}{1-\Lambda-2A}$
which is regular only for $\Lambda <(1-2A)$. The Kretschmann
invariant and Ricci scalar, are regular except for $r=0$, where
they diverge. Again the dilaton field is regular on the horizon.
The temperature and the entropy  of the black hole on the event
horizon are
\begin{eqnarray}
T &=& {1\over 4\pi} {dU \over dr}(r_{h})=\frac{1-\Lambda-2A}{4\pi N}r_{h}^{1-2N}, \\
S &=& \pi {r_{h}^{2N}}= \frac{2\pi M}{1-\Lambda-2A}r_{h}^{2N-1}.
\end{eqnarray}
which satisfy in the first law (\ref{Flaw}). In the limit $\Lambda
\rightarrow (1-2A)$, the temperature goes to zero, while the
entropy $S$ become infinite. One may note that the solution is ill
defined for $\alpha=0$. In the particular case $\alpha=1$, the
solution reduces to the (\ref{U3}), while in the absence of
Liouville potential $(\Lambda=0)$, the above solutions reduce to
(\ref{U2}).

II. $\beta=-1/\alpha$. In this case, using (\ref{Ransatz}) and
(\ref{phi}), one can easily show that eqs. (\ref{ODE1}) and
(\ref{ODE2}) have solution of the form
\begin{equation}\label{U5}
U(r)=\frac{r^{2-2N}}{N}\left(1-2A-\frac{2M}{r}+
\frac{\Lambda(1+\alpha^2)}{1-3\alpha^2} r^{2(2N-1)}\right),
\end{equation}
In order to have consistency  of all the field equations, the
electric charge satisfy in eq.(\ref{q2}). The Kretschmann
invariant and Ricci scalar, diverge at $r=0$, and both of them
vanish as $r$ goes to infinity, so there is a singularity located
at $r=0$. Note that the solution is ill defined for
$\alpha^2=1/3$. In the limit $\alpha^2\rightarrow 1$ it reduces to
the solution of section (\ref{B1}), and in the limit
$\Lambda\rightarrow 0$ the solution reduce to that with
$V(\phi)=0$.

On the other hand, if $\alpha$ and $\gamma$ go to infinity, the
solution become
\begin{equation}\label{SCH}
U(r)= 1-\frac{2M}{r}- \frac {\Lambda}{3}r^2,
\end{equation}
which is the Schwarzschild  ds/Ads  black hole, depending on the
sign of $\Lambda$. In order to investigate the causal structure of
the solution, we must investigate  the zeros of the metric
function $U(r)$. In fact, for $0<r <\infty$ the zeros of $U(r)$
are governed by the function
\begin{equation}\label{fr}
f(r) =1-2A-\frac{2M}{r}+ \frac{\Lambda(1+\alpha^2)}{(1-3\alpha^2)}
r^{2(2N-1)}.
\end{equation}

We investigate the function $g(r)=rf(r)$, for simplicity. The
cases with $\alpha^2>1/3$ and $\alpha^2<1/3$ should be considered
separately. We  should also consider the sign of the parameter
$\Lambda$ in each case.

In the first case where $\alpha^2<1/3$ and $\Lambda<0$ we may have
one horizon since $\frac{dg}{dr}>0$. But the more interesting case
happens for $\Lambda>0$ where we only obtain one local minimum at
$r=r_{min}$ where
\begin{equation}\label{rmin}
r_{min}=\left(\frac{1-2A}{\Lambda}\right)
^{\frac{\alpha^{2}+1}{2(\alpha^{2}-1)}}.
\end{equation}
The function $g(r)$ possesses horizon if $g(r_{min})< 0$. There
are two zeros for $g(r_{min})< 0$ and one degenerate zero for
$g(r_{min})=0$ which corresponds to an extremal black hole. The
condition $g(r_{min})< 0$ gives
\begin{equation}\label{nn}
M>\frac{(1-2A)(1-\alpha^{2})}{1-3\alpha^{2}}
\left(\frac{1-2A}{\Lambda}\right)
^{\frac{\alpha^{2}+1}{2(\alpha^{2}-1)}}.
\end{equation}

In the second case for $\alpha^2>1/3$ and $\Lambda<0$, the
function $g(r)$ increases monotonically, thus we can conclude that
there is one point where $g(r)=0$ which is the black hole horizon.
For $\Lambda>0$ we find local extremum for the function. The sign
of $\frac{d^{2}g(r)}{d^{2}r}$ determines whether we have local
maximum or minimum. For $\alpha^{2}>1$ we have local maximum and
$g(r_{max})$ should be positive in order to have any horizon. The
latter condition gives
\begin{equation}\label{nnn}
M<\frac{(1-2A)(1-\alpha^{2})}{1-3\alpha^{2}}
\left(\frac{1-2A}{\Lambda}\right)
^{\frac{\alpha^{2}+1}{2(\alpha^{2}-1)}}.
\end{equation}
If we have $\frac{1}{3}<\alpha^{2}<1$, then we would have local
minimum and in case of any horizon existing $g(r_{min})$ eq.
(\ref{rmin}) should be negative which implies eq. (\ref{nn}).

The above considerations show that the solutions describe black
holes with two horizons or an extremal black hole hiding a
singularity at the origin $r=0$, when the mass satisfies
(\ref{nn}), (\ref{nnn}). The radius of inner and outer horizons
can not be expressed in a closed analytical form except for the
extremal case. The radius of the extremal  solution $r_{ext}$,
coincides with $r_{min}$:
\begin{equation}
r_{ext}=\left(\frac{1-2A}{\Lambda}\right)
^{\frac{\alpha^{2}+1}{2(\alpha^{2}-1)}}=
\frac{(1-3\alpha^{2})M_{ext}}{(1-2A)(1-\alpha^{2})}.
\end{equation}

Unfortunately, because of the nature of the exponents of $r$ in
(\ref{fr}), the event horizon determined by $f(r)=0$ can not be
expressed in a closed analytical form for arbitrary $\alpha$. As
an example, we consider the  special case $\alpha=\sqrt{3}$. For
this value of $\alpha$, and large $\gamma$ limit, the action
(\ref{action}) is simply the Kaluza-Klein action which is obtained
by dimensionally reducing the five dimensional vacuum Einstein
action. For details, see \cite{frolov,GW,HH}.

In this case, there are two zeros of $f(r)$ at $r_{\pm}$:
\begin{equation}\label{alp3}
r_{\pm} = \frac{1}{\Lambda}\left(1-2A \pm \sqrt{(1-2A)^2-4\Lambda
M}\right).
\end{equation}
The extremal solution corresponds to $M_{ext}=
\frac{(1-2A)^2}{4\Lambda}$. In this case $f(r)$ has only one root
at $r_{ext}=\frac{1-2A}{\Lambda}$. When $\Lambda<
\frac{(1-2A)^2}{4M}$ we have two horizons located at $r= r_{\pm}$.
For $\Lambda>\frac{(1-2A)^2}{4M}$, there is a naked singularity at
$r=0$.

The temperature and the entropy of the black hole on the  horizons
are
\begin{eqnarray}\label{TS}
T_{\pm} &=&
\frac{{r_{\pm}}^{-3/2}}{12\pi}\left(4M+2(1-2A)r_{\pm}-3\Lambda
{r_{\pm}}^2\right),\\
S_{\pm} &=& \pi {r_{\pm}}^{3/2}.
\end{eqnarray}
which satisfy in the first law of thermodynamics.
\section{Solutions with a general  coupling  parameter  and  two  Liouville
potentials}\label{2V}
In this section, we present  exact solutions
to the EBId gravity equations with an arbitrary dilaton coupling
parameter $\alpha$ and dilaton potential
\begin{equation}\label{v2}
V(\phi) = 2\Lambda_{1} e^{2\beta_{1}\phi} +2 \Lambda_{2}
e^{2\beta_{2}\phi}.
\end{equation}
Where $\Lambda_{1}$, and $\Lambda_{2}$, $ \beta_{1}$ and $
\beta_{2}$ are  constants. This kind of  potential was previously
investigated by a number of authors both in the context of $FRW$
scalar field cosmologies \cite{ozer} and EMd black holes
\cite{CHM,yaz1}. This generalizes further the potential
(\ref{v1}). If $\beta_{1}=\beta_{2}$, then (\ref{v2}) reduces to
(\ref{v1}), so we will not repeat these solutions. Requiring
$\beta_{1}\neq \beta_{2}$, ones obtains
\begin{equation}\label{U6}
U(r)= \frac{r^{2-2N}}{N}\left(1-2A-\Lambda_{1}-\frac{2M}{r}+
\frac{\Lambda_{2}(1+\alpha^2)}{1-3\alpha^2} r^{2(2N-1)} \right).
\end{equation}
In order to fully satisfy the system of equations, the $\beta_{1}$
and $\beta_{2}$ parameters must satisfy
$\beta_{1}=1/\beta_{2}=-\alpha$, and $q$ parameter should be
satisfy in  eq. (\ref{q3}), with replacing $\Lambda \rightarrow
\Lambda_{1}$. Note that the solution is ill defined for
$\alpha^2=1/3$. In the particular case $\Lambda_{2}=0$, this
solution reduces to (\ref{U4}) and  when $\Lambda_{1}=0$, it
reduces to (\ref{U5}). The Kretschmann invariant and Ricci scalar,
diverge at $r=0$, and both of them vanish as $r$ goes to infinity,
so there is a singularity located at $r=0$. Another solution with
the same spacetime metric is generated via the discrete
transformation $\beta_{1}\longleftrightarrow \beta_{2}$ and
$\Lambda_{1}\longleftrightarrow \Lambda_{2}$.

In order to investigate the causal structure of the solution and
subsequently find the horizons (similar to what was done in the
previous section) we find the zeros of the function
\begin{equation}
f(r)=(1-2A-\Lambda_{1})-\frac{2M}{r}+\frac{\Lambda_{2}(1+\alpha^{2})}{1-3\alpha^{2}}r^{4N-2},
\end{equation}
Again, we investigate the function $g(r)=rf(r)$, for simplicity.
The cases with $\alpha^2>1/3$ and $\alpha^2<1/3$ should be
considered separately. We  should also consider the sign of the
parameter $\Lambda_{1}$  and  $\Lambda_{2}$.

For the first case, where $\alpha^2>1/3$, we certainly govern
extremum if $\Lambda_1>1/2(\Lambda_1<1/2)$ and
$\Lambda_2<0(\Lambda_2>0)$. The sign of second derivative will
show whether we have local minimum or maximum. Here, for
$1/3<\alpha^2<1(\alpha^2>1)$ and $\Lambda_2>0(\Lambda_2<0)$ the
function $f(r)$ would have local minimum and in opposite, for
$1/3<\alpha^2<1(\alpha^2>1)$ and $\Lambda_2<0(\Lambda_2>0)$ the
function $g(r)$ will have local maximum at
\begin{equation}\label{rmax}
r_{min(max)}=\left(\frac{1-2A-\Lambda_{1}}{\Lambda_{2}}\right)
^{\frac{\alpha^{2}+1}{2(\alpha^{2}-1)}}.
\end{equation}
The value of the function $g(r)$ at its extremum is
\begin{equation}\label{frmax}
g(r_{ext})=-2M+2\frac{(1-2A-\Lambda_1)(1-\alpha^2)}{1-3\alpha^2}
\left(\frac{1-2A-\Lambda_{1}}{\Lambda_{2}}\right)
^{\frac{\alpha^{2}+1}{2(\alpha^{2}-1)}}.
\end{equation}
In order to have any horizon, $g(r_{min})[g(r_{max})]$ should be
larger(less) than or equal to zero in order to possess  any local
extremum and subsequently to have any horizon for the black hole.
The case $g(r_{min})[g(r_{max})]=0$ corresponds to an extremal
black hole. The condition $g(r_{min})<0$ gives
\begin{equation}\label{namosavi}
M >\frac{(1-2A-\Lambda_1)(1-\alpha^{2})}{1-3\alpha^{2}}
\left(\frac{1-2A-\Lambda_1}{\Lambda_2}\right)
^{\frac{\alpha^{2}+1}{2(\alpha^{2}-1)}}.
\end{equation}
and we obtain the following inequality for the condition
$g(r_{max})>0$
\begin{equation}\label{namosavi}
M< \frac{(1-2A-\Lambda_1)(1-\alpha^{2})}{1-3\alpha^{2}}
\left(\frac{1-2A-\Lambda_1}{\Lambda_2}\right)
^{\frac{\alpha^{2}+1}{2(\alpha^{2}-1)}}.
\end{equation}

For the second case where $\alpha^2<1/3$ the function $g(r)$
possess local minimum for $\Lambda_2<0$ and local maximum in case
of $\Lambda_2>0$. In this case, the function diverges both at
$r=0$ and at infinity. The local minimum(maximum) happens at
(\ref{rmax}) and the value of the function $g(r_{min})$ is given
by (\ref{frmax}). Since in this case we have both a local minimum
and maximum, condition (\ref{namosavi}), should hold for both
cases.

We see that in both cases we obtain horizons for any given value
of the parameter $\alpha$. Here we express the the radius of the
extremal solution like the preceding section
\begin{equation}
r_{ext}=\left(\frac{1-2A-\Lambda_1}{\Lambda_2}\right)
^{\frac{\alpha^{2}+1}{2(\alpha^{2}-1)}}=
\frac{(1-3\alpha^{2})M_{ext}}{(1-\alpha^{2})(1-2A-\Lambda_1)}.
\end{equation}
Unfortunately, because of the nature of the exponents of $r$ in
(\ref{fr}), the event horizon determined by $g(r)=0$ can not be
expressed in a closed analytical form for arbitrary $\alpha$.

\section{Conclusion}
Born-Infeld theory and dilaton gravity are well-motivated and
extensively studied theories, not only separately, and also
coupled to each other. In this paper, we derived some classes of
exact, electrically charged, static and spherically symmetric
black hole solutions to four dimensional
Einstein-Born-Infeld-dilaton gravity without potential or with one
or two Liouville  type  potentials. The black hole solutions have
unusual asymptotics. They are neither asymptotically flat nor
asymptotically (anti-) de Sitter. In particular, in the case of
the linear electrodynamics with ${\cal L}(Y)= -{1\over 2}Y$ the
method presented here gives the well-known asymptotically non-flat
and non-(A)dS black hole solutions of the EMd  gravity \cite{CHM}.
We showed that in the presence of dilaton field,  both Kretschmann
invariant and Ricci scalar diverge at $r=0$, they remain finite
for $r\neq 0$ and tend to zero as $r\rightarrow \infty $. Thus we
have an essential singularity located at $r=0$. We found that in
the presence of Liouville type potentials for dilaton field, there
exist two classes of solutions which satisfy all the field
equations depending on  suitable choice of the $\beta$ parameter.
We also computed -for each case- temperature, entropy, charge and
mass of the black hole solutions, and find that these quantities
satisfy the first law of thermodynamics. We found that in order to
fully satisfy all the field equations consistently, there must be
a relation between the electric charge and other parameters of the
system. In general, the electric charge depends on the three
parameters  $\gamma$ , $\alpha$ and $\Lambda$.  We found that in
the large $\gamma$ and $\alpha$ limit, our solutions reduce to
Schwarzschild, and  Schwarzschild ds/Ads black holes, depending on
the sign of $\Lambda$.

Finally, it should be noted that the solutions were based on an
ansatz and consistency checks demanded  a relationship between the
parameters of the theory. An attempt for finding exact solutions
of EBId gravity by relaxing the ansatz (\ref{Ransatz}), is under
investigation. Note that the four dimensional EBId black hole
solutions obtained here are static. Therefore, it would be
interesting if one could construct charged rotating solutions of
EBId gravity in four dimensions. One can also attempt to construct
static and rotating solutions of the EBId gravity with both flat
and curved horizons in various dimensions.

\section*{Acknowledgements} We would like to thank M. H. Dehghani
for useful discussions and his comments. This work has been
supported in part by Shiraz University.


\begin{thebibliography}{99}

\bibitem{Wit1}  M. B. Green, J. H. Schwarz and E. Witten, {\it Superstring
Theory}, (Cambridge University Press, Cambridge 1987).

\bibitem{CDB1}  G. W. Gibbons and K. Maeda, Nucl. Phys. {\bf B298}, 741
(1988); T. Koikawa and M. Yoshimura, Phys. Lett. {\bf B189}, 29
(1987); D. Brill and J. Horowitz, {\em ibid.} {\bf B262}, 437
(1991).

\bibitem{CDB2}  D. Garfinkle, G. T. Horowitz and A. Strominger, Phys. Rev.
D {\bf 43}, 3140 (1991); R. Gregory and J. A. Harvey, {\em ibid.}
{\bf 47}, 2411 (1993); M. Rakhmanov, {\em ibid.} {\bf 50}, 5155
(1994).

\bibitem{MW}  S. Mignemi and D. Wiltshire, Class. Quant.
Grav. {\bf 6}, 987 (1989); D. Wiltshire, Phys. Rev. D {\bf 44},
1100 (1991); S. Mignemi and D. Wiltshire,  {\em ibid.} {\bf 46},
1475 (1992).

\bibitem{PW} S. Poletti and D. Wiltshire,  Phys. Rev. D {\bf 50}
, 7260 (1994); {\bf 52}, 3753 (1995).

\bibitem{CHM}  K. C. K. Chan, J. H. Horne and R. B. Mann, Nucl. Phys. {\bf
B447}, 441 (1995).

\bibitem{Cai1}  R. G. Cai, J. Y. Ji and K. S. Soh, Phys. Rev. D {\bf 57}, 6547
(1998); R. G. Cai and Y. Z. Zhang, {\em ibid.} {\bf 64}, 104015
(2001); R. G.  Cai and A. Wang, Phys. Rev. D {\bf 70},
084042(2004).

\bibitem{yaz1} S. S. Yazadjiev, Class.Quant. Grav. {\bf22}, 3875(2005).

\bibitem{BI} M. Born and L. Infeld, Proc. R. Soc. London  {\bf A143}, 410 (1934).

\bibitem{Fra} E. Fradkin, A. Tseytlin,  Phys. Lett. {\bf
B163}, 123 (1985); R. Matsaev, M. Rahmanov, A. Tseytlin,
 Phys. Lett. {\bf B193}, 205 (1987);E. Bergshoeff, E.
Sezgin, C. Pope, P. Townsend,  Phys. Lett. {\bf B188}, 70 (1987).

\bibitem{Cal}C. Callan, C. Lovelace, C. Nappi and S. Yost, Nucl.Phys.
{\bf B308},221(1988); O. Andreev and A. Tseytlin, Nucl.Phys.{\bf
 B311}, 221(1988); R. Leigh,  Mod. Phys. Lett . {\bf A4}, 2767 (1989).

\bibitem{wil}D. L. Wilshire,  Phys. Rev. D {\bf 38}, 2445
(1988); N. Breton {\em ibid.} {\bf 67}, 124004 (2003) ; D. A.
Rasheed, [arXiv:hep-th/9702087]; M. Aiello, R. Ferraro, G.
Giribet, [arXiv:gr-qc/0408078]; S. Fernando, D. Krug,  Gen. Rel.
Grav. {\bf35}, 129 (2003).

\bibitem{dey}Tanay Kr. Dey, Phys. Lett . {\bf B595}, 484 (2004).

\bibitem{Cai2}Rong-Gen Cai, Da-Wei Pang, Anzhong Wang, Phys.Rev. D {\bf 70},
124034 (2004).

\bibitem{YI} R. Yamazaki and D.Ida, Phys. Rev. D {\bf 64}, 024009 (2001).

\bibitem{yaz2}S. S. Yazadjiev, P. P. Fiziev, T. L. Boyadjiev, M. D.
Todorov,  Mod. Phys.Lett. {\bf A16},  2143 (2001).
 T. Tamaki and T. Torii, Phys. Rev. D {\bf 62}, 061501 (2000).
 G. Celement and D. Gal' tsov, Phys. Rev. D {\bf 62}, 124013 (2000).

\bibitem{yaz3} S. S. Yazadjiev,Phys.Rev. D {\bf 72},
044006 (2005).

\bibitem {gib}G.W. Gibbons and K. Maeda, Nucl.
Phys. {\bf B298},  741 (1988);
 D. Garfinkle, G.T. Horowitz, and A. Strominger, Phys. Rev. D {\bf 43}, 3140 (1991)
; {\em ibid.} {\bf 45}, 3888 (1992).

\bibitem{gross}  D.J. Gross and J. H. Sloan, Nucl. Phys. {\bf
B291}, 41 (1987).

\bibitem{deh1}M. H Dehghani, Iran. J. Phy.
{\bf V 5}, 69 (2005).

\bibitem{deh2}M. H Dehghani, Phys. Rev. D {\bf
71}, 064010 (2005); M. H Dehghani and N. Farhangkhah, {\em ibid.}
{\bf 71}, 044008 (2005); A. Sheykhi and N. Riazi, Int. J. Theor.
Phys. to be published, [arXiv:hep-th/0605072]; A. Sheykhi, M. H.
Dehghani, N. Riazi and J. Pakravan [arXiv:hep-th/0606237].

\bibitem {ozer}  M. \"{O}zer and M.O. Taha, Phys. Rev. D {\bf 45} (1992)
997; R. Easther,  Class. Quant.  Grav. {\bf 10}, 2203 (1993).

\bibitem{wald}R. M. Wald, \textit{General Relativity}, University
of Chicago Press, Chicago, (1984).

\bibitem{BY} J. Brown and J. York, Phys. Rev. D {\bf 47}, 1407 (1993).

\bibitem{Beck}  J. D. Beckenstein, Phys. Rev. D {\bf 7}, 2333 (1973); S. W.
Hawking, Nature (London) {\bf 248}, 30 (1974); G. W. Gibbons and
S. W. Hawking, Phys. Rev. D {\bf 15}, 2738 (1977).

\bibitem{frolov} V. Frolov, A. Zelnikov, and U. Bleyer,  Ann. Phys.
 (Leipzig) \textbf{44},  371 (1987).

\bibitem{GW} G. Gibbons and D. Wiltshire, Ann. Phys.
 (N.Y.) \textbf{167},  2061 (1986).

\bibitem{HH} J. H. Horne and G. T. Horrowitz,  Phys. Rev. D{\bf
46}, 1340 (1992).

\end{thebibliography}
\end{document}